\documentclass[12pt]{article}
\usepackage{setspace}
\usepackage{epsfig}
\usepackage{amssymb}
\def\be{\begin{equation}}
\def\ee{\end{equation}}
\def\ba{\begin{array}}
\def\ea{\end{array}}
\def\beqn{\begin{eqnarray}}
\def\eeqn{\end{eqnarray}}

\def\bt{\begin{tabular}}
\def\et{\end{tabular}}
\def\bc{\begin{center}}
\def\ec{\end{center}}
\begin{document}
\title{ Constructing the CKM and PMNS matrices from mixing data}
\author{Gulsheen Ahuja$^1$, Manmohan Gupta$^1$,
 Monika Randhawa$^2$ \\
\\
{$^1$ \it Department of Physics, Centre of Advanced Study, P.U.,
Chandigarh, India.}\\{$^2$ \it University Institute of Engineering
and Technology, P.U., Chandigarh, India.}\\{\it Email:
mmgupta@pu.ac.in}}
 \maketitle
\begin{abstract}
The CKM and PMNS matrices have been constructed based on the
latest measurements, largely free from theoretical inputs as well
as likely NP effects in the case of the former. To facilitate the
construction of the CKM matrix in the PDG representation as well
as in view of the comparatively large error in the measured value
of the CP violating phase $\delta$, the possibility of its
construction from the tree level measured CKM elements has also
been explored using the unitarity triangle. In view of the
persistent difference between the $|V_{ub}|$ exclusive and
inclusive values, we have carried out separate analyses
corresponding to these. The PMNS matrix has been constructed by
incorporating the constraints due to solar and atmospheric
neutrinos as well as by giving full variation to the Dirac-like CP
violating phase $\delta$ and considering different values of
$s_{13}$ having implications for different models of lepton mass
matrices. Taking clue from quark mixing phenomenology, an
analogous analysis of the leptonic unitarity triangle allows an
estimate of the likely presence of CP violation in the leptonic
sector.
 \end{abstract}

In the last few years, several important developments have taken
place in the context of phenomenology of Cabibbo-Kobayashi-Maskawa
(CKM) matrix \cite{ckm} as well as neutrino oscillations. On the
one hand, in the case of CKM phenomenology, we have entered the
era of precision measurements of some of the CKM parameters like $
V_{ud}, V_{us}, V_{cb}, {\rm sin}2\beta$, etc.. In this context,
based on elaborate inputs including the loop dominated parameters
such as $\epsilon_K$, $\Delta m_d$, $\Delta m_s$, $\Delta m_K$,
etc., known to be susceptible to effects of NP, several recent
updates of the detailed analyses \cite{pdgnew}-\cite{hfag} are
available which not only give insight into the CKM phenomenology
but also enable one to fix the CKM matrix. On the other hand, in
the case of neutrino oscillations, due to the solar neutrino
experiments \cite{davis}-\cite{ahmad}, atmospheric neutrino
experiments \cite{yfuk}-\cite{san} and the reactor \cite{eguchi}
as well as accelerator \cite{ahn} based experiments, there have
been continuous improvements in the corresponding neutrino mass
square differences and mixing angles, however these are not
adequate to `construct' the Pontecorvo-Maki-Nakagata-Saki (PMNS)
matrix \cite{pmnsmat} without further theoretical inputs.

The quark and neutrino mixing matrices are well known to be
related to their corresponding mass matrices which are formulated
at the GUT (Grand Unified Theories) scale, therefore a well
determined mixing matrix would have deep implications for the
GUTs. In the case of quarks, the relationship is fairly straight
forward, however the implications of elements of the mixing matrix
on those of mass matrix are not that direct. In the case of
neutrinos, an additional complication arises if these are
Majorana-like, with the corresponding mass matrices given by
seesaw mechanism \cite{seesaw}.

For a well determined mixing matrix to provide useful clues for
formulating the appropriate mass matrices, several points have to
be kept in mind. In case one uses a mixing matrix determined by
global input, then one is not sure to what extent the theoretical
inputs as well as the NP effects, in case they are present, affect
the elements of the mixing matrix, hence correspondingly one would
not be able to correlate such effects to the elements of the mass
matrices. In this approach, it would also not be an easy exercise
to correlate subtle features such as CP violating phase $\delta$
and the angles of the unitarity triangle, found from such a
matrix, to the complex nature of the elements of the mass
matrices.

To maximize the usefulness of a well determined mixing matrix one
needs to adopt a  step-wise approach for correlating the elements
of such a matrix with those of the mass matrices. In the case of
quarks, as a first step one should be able to construct the CKM
matrix which is based purely on data and on minimal number of
assumptions as well as is free from supposed effects of NP so as
to serve as a guiding stone for formulating essentials of mass
matrices. The added complications coming from subtle features of
CKM phenomenology as well as effects of NP would then be included
in the mass matrices as additional effects.

A data based CKM mixing matrix can be determined very well in case
one has a knowledge of the CP violating phase $\delta$ along with
the mixing elements $V_{us}$, $V_{cb}$ and $|V_{ub}|$ which, in
the PDG representation are directly related to the three
respective mixing angles. The elements $V_{us}$ and $V_{cb}$ are
very well determined \cite{pdgnew}, whereas the CP violating phase
$\delta$ is known with comparatively large error bars
\cite{pdgnew,cfitter,glw}. In the case of $|V_{ub}|$, the
exclusive and inclusive measurements have yet to approach a common
value, therefore it is desirable \cite{ali} to carry out a
separate analysis pertaining to these, as has also been done very
recently by UTfit Collaboration \cite{utfitnew}. In the case of
neutrinos, the mixing matrix should preferably be determined from
oscillation data, however in the absence of availability of
adequate data pertaining to one of the mixing angles one needs to
include some input from theoretical models as well.

The purpose of the present paper is the construction of the CKM
and PMNS matrices based on the latest measurements, largely free
from theoretical inputs as well as likely NP effects in the case
of the former. Further, in view of the comparatively large error
bars associated with the measured CP violating phase $\delta$ as
well as its importance in constructing the matrix, we intend to
explore the possibility of its construction from the tree level
measured CKM elements. Furthermore, employing the unitarity based
parametrization as well as measured values of $V_{us}$, $V_{cb}$
and the CP violating phase $\delta$ we would like to construct the
CKM matrix for both $|V_{ub}|$ exclusive and inclusive values. For
the PMNS matrix, we have attempted to construct the matrix using
the unitarity based PDG parametrization and presently available
data from oscillation experiments. Further, we have attempted to
explore the likely range of Dirac-like CP violating phase $\delta$
using $J_l$, the Jarlskog's rephasing invariant parameter in the
leptonic sector .

Most of the present day analyses, related to CKM phenomenology,
have been carried out using the Wolfenstein-Buras parametrization
\cite{wolfbur} of the mixing matrix, however, in the present case
we find PDG representation \cite{pdgnew} to be more useful as the
three angles are directly related to measured CKM elements.
Similarly, in the case of neutrino mixing matrix, we adopt the PDG
representation. Also, it may be noted that in the PDG
representation the unitarity is built-in, however for finding the
angles $s_{12}$, $s_{23}$, $s_{13}$ we use the measured values of
the elements $V_{us}$, $V_{cb}$, $|V_{ub}|$, without invoking the
unitarity constraints on these. For ready reference as well as to
facilitate discussion of results, we begin with the quark mixing
phenomenon, often expressed in terms of a $3 \times 3$ quark
mixing matrix $V_{CKM}$ as
 \be \left( \ba {c} d~^{\prime} \\ s~^{\prime} \\ b~^{\prime} \ea \right)
  = \left( \ba{ccc} V_{ud} & V_{us} & V_{ub} \\ V_{cd} & V_{cs} &
  V_{cb} \\ V_{td} & V_{ts} & V_{tb} \ea \right)
 \left( \ba {c} d\\ s \\b \ea \right),  \label{qm}  \ee
which in the PDG representation \cite{pdgnew} involving angles $\theta_{12},
\theta_{23}, \theta_{13}$ and the phase $\delta$ is
given as
   \be V_{CKM}=\left( \ba{ccl} c_{12} c_{13} & s_{12} c_{13} &
  s_{13}e^{-i \delta} \\ - s_{12} c_{23} - c_{12} s_{23}
  s_{13} e^{i \delta} & c_{12} c_{23} - s_{12} s_{23}
  s_{13} e^{i \delta} & s_{23} c_{13} \\ s_{12} s_{23} - c_{12}
  c_{23} s_{13} e^{i \delta} & - c_{12} s_{23} - s_{12} c_{23}
  s_{13} e^{i \delta} & c_{23} c_{13} \ea \right), \label{qmm} \ee
 with $c_{ij}= {\rm cos}~ \theta_{ij}$ and $s_{ij}= {\rm sin}~
\theta_{ij}$ for $i,j=1,2,3$. In this representation, noting
$c_{13}\cong1$, one can consider $V_{us} \cong s_{12}$, $V_{cb}
\cong s_{23}$ and $|V_{ub}| \cong s_{13}$, henceforth $|V_{ub}|$
would be written as $V_{ub}$.

In a similar manner, neutrino mixing PMNS matrix \cite{pmnsmat} is
given by \be \left( \ba{c} \nu_e \\ \nu_{\mu} \\ \nu_{\tau} \ea
\right)
  = \left( \ba{ccc} U_{e1} & U_{e2} & U_{e3} \\ U_{\mu 1} & U_{\mu 2} &
  U_{\mu 3} \\ U_{\tau 1} & U_{\tau 2} & U_{\tau 3} \ea \right)
 \left( \ba {c} \nu_1\\ \nu_2 \\ \nu_3 \ea \right),  \label{nm}  \ee
where $ \nu_{e}$, $ \nu_{\mu}$, $ \nu_{\tau}$ are the flavor
eigenstates and $ \nu_1$, $ \nu_2$, $ \nu_3$ are the mass
eigenstates. Following PDG representation analogous to the quark
mixing case, involving three angles and the Dirac-like CP
violating phase $\delta$ as well as the two Majorana phases
$\alpha_1$, $\alpha_2$, the PMNS matrix $U_{PMNS}$ can be written
as, \beqn U_{PMNS}={\left( \ba{ccl} c_{12} c_{13} & s_{12} c_{13}
& s_{13}e^{-i \delta} \\ - s_{12} c_{23} - c_{12} s_{23} s_{13}
e^{i \delta} & c_{12} c_{23} - s_{12} s_{23} s_{13} e^{i \delta} &
s_{23} c_{13}
\\ s_{12} s_{23} - c_{12} c_{23} s_{13} e^{i \delta} & - c_{12}
s_{23} - s_{12} c_{23} s_{13} e^{i \delta} & c_{23} c_{13} \ea
\right)} \left( \ba{ccc} e^{i \alpha_1/2} & 0 & 0 \\ 0 &e^{i
\alpha_2/2} & 0 \\ 0 & 0  & 1 \ea \right). \label{nmm} \eeqn The
Majorana phases $\alpha_1$ and $\alpha_2$ do not play any role in
neutrino oscillations and henceforth would be dropped from the
discussion.

In the PDG representation, as has already been mentioned, the CKM
matrix can immediately be constructed in case $V_{us}$, $V_{cb}$,
$V_{ub}$ as well as the CP violating phase $\delta$ are well
measured. However, in view of the large error bars associated with
the phase $\delta$, as a first step towards our analysis, we
explore the possibility of its construction from the CKM elements
measured at tree level by considering the unitarity triangle which
involves these elements. We begin by considering the constraints
due to unitarity of the CKM matrix, defined as \be
\sum_{\alpha=d,s,b} V_{i \alpha} {V^*_{j \alpha}} =\delta_{ij},
\label{ut1} \ee
 \be \sum_{i=u,c,t} V_{i \alpha}
{V^*_{i \beta}} =\delta_{\alpha\beta}, \label{ut2} \ee where Greek
indices run over the down type quarks $(d,s,b)$ and Latin ones run
over the up type quarks $(u,c,t)$. Unitarity implies nine
relations, three in terms of normalization conditions, given by
equation (\ref{ut1}), sometimes also referred to as `weak
unitarity conditions' and the six non-diagonal relations, given by
equation (\ref{ut2}), also expressed through the six unitarity
triangles in the complex plane. Because of the strong hierarchical
nature of the CKM matrix elements as well as the limitations
imposed by the present level of measurements, it is difficult to
study the implications of normalization relations, therefore, the
six relations implied by equation (\ref{ut2}) are used to study
the implications of unitarity on CKM phenomenology. The usual
unitarity triangle, also referred to as the $db$ triangle, is
expressed through the relation
  \be V_{ud} {V_{ub}}^* + V_{cd} {V_{cb}}^* + V_{td}
{V_{tb}}^* =0\,.\label{db} \ee Using the above expression and by
replacing $b$ by $s$ we get the relation for the $ds$ triangle and
by replacing $d$ by $s$ we get the relation for the $sb$ triangle.
In a similar manner involving the orthogonality of the rows of the
CKM matrix, we get the relations for the other three triangles,
e.g., rows I and II give the expression for $uc$ triangle, rows I
and III give the $ut$ triangle and rows II and III give the $ct$
triangle.

It may be noted that the well known $db$ triangle involves some
elements which have not been measured directly at the tree level.
In fact, only the $uc$ triangle defined as
 \be V_{ud} {V_{cd}}^* + V_{us} {V_{cs}}^* +
V_{ub} {V_{cb}}^* =0\,,\label{uc} \ee has sides which involve CKM
elements measured at the tree level. It may be noted that this
triangle is highly skewed, with one side being very small as
compared to the other two, therefore extracting information about
the CP violating phase $\delta$ from this triangle is a non
trivial job \cite{mon}. Following the procedure given in
\cite{mon}, the phase $\delta$ can be calculated using the
Jarlskog's rephasing invariant parameter $J$, equal to twice the
area of any of the unitarity triangle, through the relation
\be
J=s_{12}s_{23}s_{13}c_{12}c_{23}c_{13}^2 \,{\rm sin}\,\delta.
\label{jd}\ee

\begin{table}
\begin{tabular}{|l|l|}  \hline
Parameter & PDG 2006 values \cite{pdgnew}
\\ \hline$V_{ud}$ & 0.97377 $\pm$ 0.00027 \\
$V_{us}$ & 0.2257 $\pm$ 0.0021 \\
 $V_{cd}$ & 0.230 $\pm$ 0.011\\
 $V_{cs}$ & 0.957 $\pm$ 0.017 $\pm$ 0.093\\
$V_{cb}$ & (41.6 $\pm$ 0.6) $\times 10^{-3}$ \\
 $V_{ub}$~(excl.) & (3.84 $^{+0.67}_{-0.49}) \times 10^{-3}$\\
 $V_{ub}$~(incl.) & (4.40 $\pm$ 0.20 $\pm$ 0.27) $\times 10^{-3}$ \\
 $\delta$ & (63.0$^{+15.0}_{-12.0})^{\rm o}$\\
 \hline
\end{tabular}
\caption{PDG 2006 values of some of the CKM parameters used in the
present analysis.} \label{qinput}
\end{table}

In table \ref{qinput}, we present the PDG 2006 values
\cite{pdgnew} of some of the CKM elements and the CP violating
phase $\delta$. For the $uc$ triangle, assuming Gaussian
probability density distribution for the CKM matrix elements, the
distribution of $J$, shown in figure \ref{jhist} for the exclusive
value of $V_{ub}$, is obtained by Monte Carlo simulations. From
this figure we obtain
\be
J= (2.865 \pm 0.968) \times 10^{-5},\ee from a similar figure, not
presented here, for the inclusive value of $V_{ub}$ we can get
\be
J= (3.203 \pm 0.943) \times 10^{-5}.\ee It may be mentioned that
in the figures we have considered only those points for which
$J\neq0$. The large error bars on above mentioned values are
essentially due to the fact that the number of points where
$J\neq0$ is much less than the total number of possible points.
The above values of $J$ have been found by fitting a Gaussian to
the respective figures. Interestingly, we find that the above
mentioned $J$ values are inclusive of that found by PDG group
through their recent global analysis \cite{pdgnew}. The values of
$\delta$ corresponding to exclusive and inclusive $V_{ub}$ found
through equation (\ref{jd}) are also compatible with the
experimentally determined $\delta$ given by PDG 2006. However, it
seems that the present level of accuracy of the tree level
measured CKM elements is not enough to find $\delta$ with a better
accuracy as compared to its measured value, therefore its value
given by PDG 2006 is being used for constructing the present CKM
matrix.

The CKM matrix for the exclusive as well as inclusive values of
$V_{ub}$ can now be easily constructed using the experimentally
determined elements $V_{us}$, $V_{cb}$ and the measured CP
violating phase $\delta$. For the exclusive value of $V_{ub}$ with
inputs at 1$\sigma$ C.L., the matrix is as follows,
 \be V_{CKM} = \left( \ba{ccc}
  0.9736 ~{\rm to}~ 0.9747 &   0.2236~ {\rm to} ~0.2282 &  0.0035 ~{\rm to} ~0.0041 \\
 0.2234 ~{\rm to}~ 0.2280  &   0.9728 ~{\rm to}~ 0.9738 &  0.0409~ {\rm to}~ 0.0423\\
0.0074 ~{\rm to}~ 0.0093  &  0.0402 ~{\rm to}~ 0.0416 &  0.9990~
{\rm to}~ 0.9991 \ea \right). \label{matexp} \ee The matrix for
the inclusive value of $V_{ub}$ as well as the average value of
$V_{ub}$, considered by PDG 2006, differs from the above matrix
only in the respective $V_{ub}$ values being 0.0043 to 0.0047 and
0.0040 to 0.0046. It may be noted that $V_{ub}\ll V_{cb}\ll
V_{us}$ ensures that refinements in $V_{ub}$ would hardly affect
the rest of the elements of the CKM matrix as well as our
conclusions in this regard.

It needs to be emphasized that this matrix has been constructed by
using the unitarity based PDG parametrization, however without
incorporating the full constraints due to unitarity, for a
detailed discussion regarding this we refer the reader to
\cite{uni}. Since we have used only the experimental values of
$V_{us}$, $V_{cb}$, $V_{ub}$ and the CP violating phase $\delta$,
this matrix is supposedly free from effects of NP. A comparison of
its elements with their predictions from models of mass matrices
based on GUTs or measurements in different experiments could
provide clues to the possibility of NP. The matrix constructed
above is fully compatible with the matrix constructed by PDG group
using global inputs. In particular, we find that the PDG mean
value of the element $V_{td}$, sensitive to both loop and NP
effects, is very near to the present mean value. This value,
however, looks to be somewhat higher than the `experimentally
found' $V_{td}$ \cite{pdgnew, vtd}, suggesting the need for
further experimental scrutiny in this case.

In the context of mass matrices, such a matrix would provide the
possibility of deciphering the appropriate textures which are in
principal agreement with the data. This can be achieved by either
finding the elements $V_{us}$, $V_{cb}$, $V_{ub}$ as well as the
phase $\delta$ or by finding the entire matrix which should then
be compared with the mixing matrix given in equation
(\ref{matexp}). It may also be added that such an exercise would
enable one to find the phase structure of mass matrices which are
in broad agreement with the data, hence would throw light on the
mechanism as well as NP effects affecting the generation of phases
at the GUT scale.

Coming to the case of neutrinos, adopting the three neutrino
framework, several authors have presented updated information
regarding the neutrino mass and mixing parameters obtained by
carrying out detailed global analyses \cite{recana}. The results
from solar and atmospheric neutrino experiments allow one to
respectively determine the two mixing angles $s_{12}$ and $s_{23}$
rather well. However, situation regarding the angle $s_{13}$ is
not well defined, at present only its upper bound \cite{chooz} is
available. The latest situation regarding masses and mixing angles
at 3$\sigma$ C.L. is summarized as follows \cite{recana},
\be
 \Delta m_{12}^{2} = (7.1 - 8.9)\times
 10^{-5}~\rm{eV}^{2},~~~~
 \Delta m_{23}^{2} = (2.0 - 3.2)\times 10^{-3}~ \rm{eV}^{2},
 \label{solatmmass}\ee
\be
{\rm sin}^2\,\theta_{12}  =  0.24 - 0.40,~~~
 {\rm sin}^2\,\theta_{23}  =  0.34 - 0.68,~~~
 {\rm sin}^2\,\theta_{13} \leq  0.040. \label{s13}
\ee

For the construction of the PMNS matrix, we have adopted the PDG
parametrization which satisfies unitarity by construction,
however, we are not constraining the output further by imposing
any of the nine unitarity relations, some of which might be
amenable to experimental data. It may be noted that in the absence
of any definite clues about $s_{13}$, we have considered those
values which have theoretical implications. It is very well
recognized that the value of $s_{13}$ would have deep implications
for the neutrino oscillation phenomenology, in particular, very
recently, Albright {\it et al.} \cite{albright} have carried out a
very detailed and exhaustive analysis wherein they have studied
the implications of the values of $s_{13}$ on various leptonic and
grand unified models of neutrino masses and mixings. Needless to
say that the value of $s_{13}$ would not only have implications
for theoretical ideas such as quark-lepton complementarity, etc.
\cite{mnsm}-\cite{bk} but would also determine future direction of
neutrino phenomenology \cite{white}-\cite{minakata}. Keeping this
in mind, we have chosen a few representative values which cover
most of these attempts.

Broadly speaking, theoretical implications result in $s_{13}$
taking values around 0.05, 0.10 and 0.15. Therefore, we have
constructed the PMNS matrix by taking above values of $s_{13}$ and
attaching 20$\%$ errors along with these. For appropriate
construction of the PMNS matrix, we have made use of the 3$\sigma$
C.L. range of the two well determined mixing angles, $s_{12}$ and
$s_{23}$, however for $s_{13}$, we have considered the ranges
$s_{13}$=(i) 0.05 $\pm$ 0.01 , (ii) 0.1 $\pm$ 0.02 and (iii) 0.15
$\pm$ 0.03. By giving full variation to the phase $\delta$, we
obtain the following mixing matrices corresponding to the above
mentioned $s_{13}$ ranges,

(i) \be U_{PMNS} = \left(  \ba{ccc}
  0.7610~ {\rm to}~ 0.8850 &   0.4893~ {\rm to}~ 0.6217 &  0.04~ {\rm to}~0.06\\
 0.2745 ~{\rm to}~ 0.4942  &   0.4460 ~{\rm to}~ 0.6984  &  0.5781~ {\rm to}~ 0.8111\\
 0.2820 ~{\rm to}~ 0.4885 & 0.4521 ~{\rm to}~ 0.6919 & 0.5695~{\rm to}~
 0.8195
 \ea \right), \label{r1} \ee

(ii) \be U_{PMNS} = \left(  \ba{ccc}
  0.7575~ {\rm to}~ 0.8819 &   0.4871~ {\rm to}~ 0.6193 &  0.08~ {\rm to}~0.12\\
 0.2414 ~{\rm to}~ 0.5200  &   0.4242 ~{\rm to}~ 0.7148  &  0.5756~ {\rm to}~ 0.8079\\
 0.2478 ~{\rm to}~ 0.5133 & 0.4296 ~{\rm to}~ 0.7085 & 0.5674~{\rm to}~
 0.8164
 \ea \right), \label{r2} \ee

(iii) \be U_{PMNS} = \left(\ba{ccc}
  0.7527~ {\rm to}~ 0.8761 & 0.4839~ {\rm to}~ 0.6152 & 0.12~ {\rm to}~0.15\\
 0.2086 ~{\rm to}~ 0.5468  & 0.4018 ~{\rm to}~ 0.7312  & 0.5718~ {\rm to}~0.8027\\
 0.2130 ~{\rm to}~ 0.5389 & 0.4069 ~{\rm to}~ 0.7256 & 0.5634~{\rm to}~
 0.8110
 \ea \right). \label{r3} \ee
In the case of quarks $s_{13} \ll s_{23} \ll s_{12}$, therefore
the elements $V_{cd}$, $V_{cs}$, $V_{td}$, $V_{ts}$ of the CKM
matrix are hardly affected by it, however in the case of
neutrinos, although $s_{13}$ is small as compared to the other two
angles yet it is not as small as in the case of quarks, therefore
we find that changes in $s_{13}$ fairly affect the corresponding
above mentioned elements of the PMNS matrix. This observation
would have implications for the predictions of various models of
neutrino mass matrices. It is interesting to note that the present
neutrino mixing matrices are fully compatible with a very recent
construction of a mixing matrix by Bjorken {\it et al.}
\cite{bjorken} assuming democratic trimaximally mixed $\nu_2$ mass
eigenstate as well as with the one presented by Giunti
\cite{giunti}.

One of the key issue in the context of neutrino oscillations is
the possibility of observing CP violation in the leptonic sector.
In this context, an important parameter in the leptonic sector,
parallel to quark mixing case, is the Jarlskog's rephasing
invariant parameter $J_l$. In most of the recent analyses of
neutrino oscillation phenomenology \cite{fukugita,xing}, it is
usual to calculate the upper limit of $J_l$, however, in the
present analysis, following the quark mixing analogy, we attempt
to calculate the most probable range of $J_l$ keeping in mind the
present uncertainties regarding the mixing angle $s_{13}$ and the
Dirac-like CP violating phase $\delta$. To this end, considering
each element to be Gaussian and following the same procedure as in
the quark case, using the first two rows of the matrix given in
equation (\ref{r2}), we obtain $J_l$ as \be J_l= 0.0192 \pm
0.0079, \label{jval}\ee the distribution of $J_l$ has been plotted
in figure \ref{jnu}. It may be noted that similar calculations can
also be carried out using the matrices given in equations
(\ref{r1}) and (\ref{r3}), however, the values of $J_l$ are not
much different from the one given in equation (\ref{jval}). Also,
we can calculate $J_l$ using other triangles as well, yielding
more or less the same value. Since the input values have been
taken at 3$\sigma$ C.L., therefore the calculated value of $J_l$
can be considered a fairly reasonable estimate of likely CP
violation in the leptonic sector. We find the corresponding
$\delta$ in the range $15^{\rm o} - 165^{\rm o}$. Although it
cannot be predicted as a signal for CP violation, yet it does
suggests that in case $s_{13}\sim0.1$, CP violation in the
leptonic sector may be there.

To summarize, we have carried out the construction of the CKM and
PMNS matrices based on the latest measurements, largely free from
theoretical inputs as well as likely NP effects in the case of the
former. In view of the comparatively large error bars associated
with the measured CP violating phase $\delta$, we have first
explored the possibility of its construction from the tree level
measured CKM elements. In the context of CKM matrix, employing the
unitarity based parametrization, we have constructed the matrix by
using the precisely measured $V_{us}$, $V_{cb}$, the CP violating
phase $\delta$ found from data as well as $V_{ub}$ exclusive and
inclusive values.

The same procedure is followed for the construction of the PMNS
matrix, however, in this case we have carried out the construction
for different values of $s_{13}$, having implications for
different models of lepton mass matrices, as well as by giving
full variation to the CP violating phase $\delta$. The Jarlskog's
rephasing invariant parameter in the leptonic sector $J_l$ has
been calculated by finding the area of the leptonic unitarity
triangle, its comparatively large value suggests that CP violation
may be there in the leptonic sector.

  \vskip 0.2cm
{\bf Acknowledgements} \\ The authors would like to thank S.D.
Sharma and Sanjeev Kumar for useful discussions. G.A. and M.G.
would like to thank DAE, BRNS (grant No.2005/37/4/BRNS), India,
for financial support. M.R. would like to thank the Director, UIET
for providing facilities to work.

\end{document}